\documentclass[preprint,aps,12pt,showpacs,nofootinbib,tightenlines]{revtex4}
\usepackage{amsmath}
\usepackage{amssymb}
\usepackage{CJK}
\usepackage{epsfig}
\usepackage{graphicx}
\usepackage{subfigure}
\usepackage{slashed}
\begin{document}
\begin{CJK*}{GBK}{song}
\def\pslash{\rlap{\hspace{0.02cm}/}{p}}
\def\eslash{\rlap{\hspace{0.02cm}/}{e}}
\title {Rare Higgs three body decay induced by top-Higgs FCNC coupling
in the littlest Higgs Model with T-parity}
\author{Bing-Fang Yang}\email{yangbingfang@htu.edu.cn}
\author{Zhi-Yong Liu}\email{021168@htu.cn}
\author{Ning Liu}\email{wlln@mail.ustc.edu.cn}
\affiliation{ $^1$College of Physics $\&$ Electronic Engineering,
Henan Normal University, Xinxiang 453007, China
   \vspace*{1.5cm}  }

\begin{abstract}

Motivated by the search for flavor-changing neutral current (FCNC)
top quark decays at the LHC, we calculate rare Higgs three body
decay $H\rightarrow Wbc$ induced by top-Higgs FCNC coupling in the
littlest Higgs model with T-parity(LHT). We find that the branching
ratios of $H\rightarrow Wbc$ in the LHT model can reach $\mathcal
O(10^{-7})$ in the allowed parameter space.
\end{abstract}
 \maketitle
\section{ Introduction}
\noindent

The discovery of a Higgs-like resonance near 125 GeV
\cite{LHC-Higgs} at the LHC is a great triumph for theoretical and
experimental particle physics. So far, most measurements of this new
particle are consistent with the Standard Model (SM) prediction, but
the experimental investigation of this new particle has only just
begun. It is not impossible that more in-depth studies will reveal
its non-SM properties.

Compared with the normal decay modes, the flavour-changing neutral
current (FCNC) decays are highly suppressed in the SM due to the
Glashow-Iliopoulos-Maiani (GIM) mechanism\cite{GIM}. So, any large
enhancements in these branching ratios will be smoking-gun signals
beyond the SM.

As the heaviest known elementary particle, the top quark is widely
speculated to be sensitive to the electroweak symmetry breaking
(EWSB) mechanism and new physics at TeV-scale. An interesting
possibility is the presence of FCNC interactions between the Higgs
boson and the top quark. This interaction not only participate in
the top quark FCNC decays\cite{topfcnc}, but also participate in the
Higgs FCNC decays\cite{hfcnc}.

Except for the dominant decay mode $H\rightarrow b\bar{b}$, the so
called below-threshold decay modes induced by the $HVV(V = W;Z)$
couplings are also very important, where the decay $H\rightarrow VV$
with one (or two) V's being off-shell and decaying to fermions. In
some new physics, the decay mode of Higgs bosons is much richer and
3-body decays may be even more important. Now, almost all Higgs
boson decay modes have been measured at the LHC, but they are
plagued by large SM backgrounds. So, the rare Higgs 3-body decays
may bring us more surprises. In some new physics models, the GIM
suppression can be relaxed and/or new particles can contribute to
the loops, so that the top-Higgs FCNC couplings $tqH$, especially
the $tcH$ coupling, can be enhanced by orders of magnitude larger
than those of the SM\cite{topfcncNP}.

In this paper, we study the rare Higgs 3-body decay $H\rightarrow
Wbc$ induced by top-Higgs FCNC coupling in the littlest Higgs Model
with T-parity(LHT). This decay includes the FCNC vertex $tcH$, which
receives the contribution from the new T-odd gauge bosons and T-odd
fermions. The results of this process will help to test the SM and
probe the LHT model.

The paper is organized as follows. In Sec.II we give a brief review
of the LHT model related to our work. In Secs.III we calculate the
rare Higgs 3-body decay $H\rightarrow Wbc$ induced by top-Higgs FCNC
coupling in unitary gauge under current constraints. Finally, we
draw our conclusions in Sec.IV.

\section{ A brief review of the LHT model}
\noindent The LHT model is based on an $SU(5)/SO(5)$ non-linear
$\sigma$ model\cite{LHT}. At the scale $f\sim \mathcal O$ (TeV), the
$SU(5)$ global symmetry is broken down to $SO(5)$ by the vacuum
expectation value (VEV) of the $\sigma$ field, $\Sigma_0$, given by
\begin{eqnarray}
\Sigma_0=\langle\Sigma\rangle
\begin{pmatrix}
{\bf 0}_{2\times2} & 0 & {\bf 1}_{2\times2} \\
                         0 & 1 &0 \\
                         {\bf 1}_{2\times2} & 0 & {\bf 0}_{2\times 2}
\end{pmatrix}.
\end{eqnarray}
After the global symmetry is broken, there arise 14 Goldstone
bosons(GB) which are described by the ``pion'' matrix $\Pi$. Then
the kinetic term for the GB matrix can be expressed in the standard
non-linear sigma model formalism as
\begin{equation}
\Sigma = e^{i\Pi/f}\ \Sigma_0\ e^{i\Pi^T/f}\equiv e^{2i\Pi/f}\
\Sigma_0. \label{sigmaA}
\end{equation}

The $\sigma$ field kinetic Lagrangian is given by
\begin{equation}
\mathcal{L}_{\rm K}= \frac{f^{2}}{8} {\rm Tr} | D_{\mu} \Sigma |^2,
\label{kinlag}
\end{equation}
with the $[SU(2)\otimes U(1)]^2$ covariant derivative defined by
\begin{equation}
D_{\mu} \Sigma = \partial_{\mu} \Sigma - i \sum_{j=1}^2 \left[ g_{j}
W_{j\,\mu}^{a} (Q_{j}^{a}\Sigma + \Sigma Q_{j}^{a\,T}) + g'_{j}
B_{j\,\mu} (Y_{j} \Sigma+\Sigma Y_{j}^{T}) \right],
\end{equation}
where $W_{j}^\mu = \sum_{a=1}^{3} W_{j}^{\mu \, a} Q_{j}^{a}$ and
$B_{j}^\mu = B_{j}^{\mu} Y_{j}$ are the heavy $SU(2)$ and $U(1)$
gauge bosons, with $Q_j^a$ and $Y_j$ the gauge generators, $g_j$ and
$g'_j$ are the respective gauge couplings.

The VEV $\Sigma_0$ also breaks the gauged subgroup $\left[
SU(2)\times U(1) \right]^2$ of the $SU(5)$ down to the SM
electroweak $SU(2)_L \times U(1)_Y$. At $\mathcal O(v^{2}/f^{2})$ in
the expansion of the Lagrangian (\ref{kinlag}), the masses of the
T-parity partners of the $W$ boson ($W_{H}^{\pm}$), $Z$ boson
($Z_{H}$) and photon ($A_{H}$) after EWSB are given by
\begin {equation}
M_{W_{H}}=M_{Z_{H}}=gf(1-\frac{v^{2}}{8f^{2}}),~~M_{A_{H}}=\frac{g'f}{\sqrt{5}}
(1-\frac{5v^{2}}{8f^{2}})
\end {equation}
where $g$ and $g'$ denote the SM $SU(2)$ and $U(1)$ gauge couplings,
respectively. $v$ represents the VEV of the Higgs doublet, which is
related to the SM Higgs VEV $v_{SM} = 246$GeV through the following
formula:
\begin{equation}
v = \frac{f}{\sqrt{2}} \arccos{\left( 1 -
\frac{v_\textrm{SM}^2}{f^2} \right)} \simeq v_\textrm{SM} \left( 1 +
\frac{1}{12} \frac{v_\textrm{SM}^2}{f^2} \right).
\end{equation}

In the quark sector, the T-odd mirror partners for each SM quark are
added to preserve the T-parity. The up and down-type mirror quarks
can be denoted by $u_{H}^{i}$ and $d_{H}^{i}$, where $i$($=1,2,3$)
is the generation index. One can write down a Yukawa interaction to
give masses to the mirror quarks
\begin{eqnarray}
\mathcal{L}_{\rm mirror}=-\kappa_{ij}f\left(\bar\Psi_2^i\xi +
  \bar\Psi_1^i\Sigma_0\Omega\xi^\dagger\Omega\right)\Psi_R^j+h.c.\
\end{eqnarray}
After the EWSB, their masses up to $\mathcal O(v^{2}/f^{2})$ are
given by
\begin{equation}
m_{d_{H}^{i}}=\sqrt{2}\kappa_if, ~~m_{u_{H}^{i}}=
m_{d_{H}^{i}}(1-\frac{v^2}{8f^2})
\end{equation}
where $\kappa_i$ are the eigenvalues of the mass matrix $\kappa$.

Under T -parity, in order to cancel the large radiative correction
to Higgs mass parameter induced by top quark, an additional T-even
heavy quark $T^{+}$ and its T-odd mirror partner $T^{-}$ are
introduced. Their masses are given by
\begin{eqnarray}
m_{T^{+}}&=&\frac{f}{v}\frac{m_{t}}{\sqrt{x_{L}(1-x_{L})}}[1+\frac{v^{2}}{f^{2}}(\frac{1}{3}-x_{L}(1-x_{L}))]\\
m_{T^{-}}&=&\frac{f}{v}\frac{m_{t}}{\sqrt{x_{L}}}[1+\frac{v^{2}}{f^{2}}(\frac{1}{3}-\frac{1}{2}x_{L}(1-x_{L}))]
\end{eqnarray}
where $x_{L}$ is the mixing parameter between the top-quark and
heavy quark $T^{+}$. This mixing parameter can also be expressed by
a ratio $R=\lambda_{1}/\lambda_{2}$ with
\begin{equation}
x_{L}=\frac{R^{2}}{1+R^{2}}
\end{equation}
where $\lambda_{1}$ and $\lambda_{2}$ are two dimensionless top
quark Yukawa couplings.

When the mass matrix $\sqrt{2}\kappa_{ij}f$ is diagonalized by two
$U(3)$ matrices, a new flavor structure can come from the mirror
fermions. In the mirror quark sector, the existence of two CKM-like
unitary mixing matrices $V_{Hu}$ and $V_{Hd}$ is one of the
important ingredients. It's worth noting that $V_{Hu}$ and $V_{Hd}$
are related through the SM CKM matrix:
\begin{equation}
V_{Hu}^{\dag}V_{Hd}=V_{CKM}.
\end{equation}

Follow Ref.\cite{vhd}, the matrix $V_{Hd}$ can be parameterized with
three angles $\theta^d_{12},\theta^d_{23},\theta^d_{13}$ and three
phases $\delta^d_{12},\delta^d_{23},\delta^d_{13}$
\begin{eqnarray}
V_{Hd}=
\begin{pmatrix}
c^d_{12}c^d_{13}&s^d_{12}c^d_{13}e^{-i\delta^d_{12}}&s^d_{13}e^{-i\delta^d_{13}}\\
-s^d_{12}c^d_{23}e^{i\delta^d_{12}}-c^d_{12}s^d_{23}s^d_{13}e^{i(\delta^d_{13}-\delta^d_{23})}&
c^d_{12}c^d_{23}-s^d_{12}s^d_{23}s^d_{13}e^{i(\delta^d_{13}-\delta^d_{12}-\delta^d_{23})}&
s^d_{23}c^d_{13}e^{-i\delta^d_{23}}\\
s^d_{12}s^d_{23}e^{i(\delta^d_{12}+\delta^d_{23})}-c^d_{12}c^d_{23}s^d_{13}e^{i\delta^d_{13}}&
-c^d_{12}s^d_{23}e^{i\delta^d_{23}}-s^d_{12}c^d_{23}s^d_{13}e^{i(\delta^d_{13}-\delta^d_{12})}&
c^d_{23}c^d_{13}
\end{pmatrix}
\end{eqnarray}

For the down-type quarks and charged leptons, there are two possible
ways to construct the Yukawa interaction, which are denoted as Case
A and Case B\cite{caseAB}. At order $\mathcal{O} \left( v_{SM}^4/f^4
\right)$, the corresponding corrections to the Higgs couplings are
given by ($d \equiv d,s,b,l^{\pm}_i$)
\begin{eqnarray}
    \frac{g_{h \bar{d} d}}{g_{h \bar{d} d}^{SM}} &=& 1-
        \frac{1}{4} \frac{v_{SM}^{2}}{f^{2}} + \frac{7}{32}
        \frac{v_{SM}^{4}}{f^{4}} \qquad \text{Case A} \nonumber \\
    \frac{g_{h \bar{d} d}}{g_{h \bar{d} d}^{SM}} &=& 1-
        \frac{5}{4} \frac{v_{SM}^{2}}{f^{2}} - \frac{17}{32}
        \frac{v_{SM}^{4}}{f^{4}} \qquad \text{Case B}
    \label{dcoupling}
\end{eqnarray}

\section{Branching ratio for $H\rightarrow Wbc$ in the LHT model}

\noindent

\begin{figure}[htbp]
\scalebox{0.45}{\epsfig{file=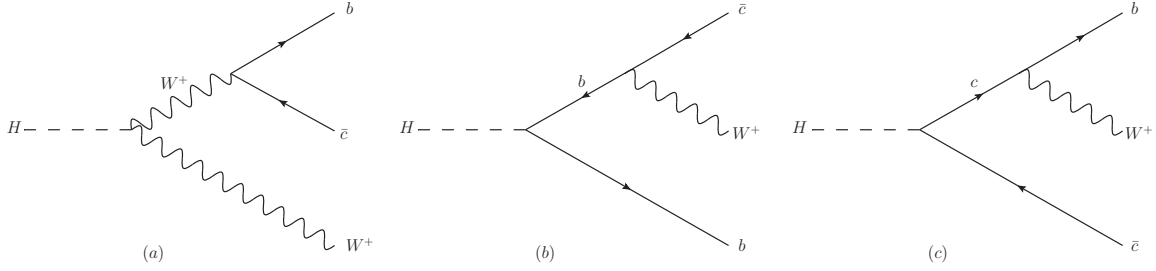}} \caption{Feynman diagrams
of the decay $H\rightarrow W^{+}b\bar{c}$ at tree
level.}\label{hwbc2}
\end{figure}

\begin{figure}[htbp]
\scalebox{0.45}{\epsfig{file=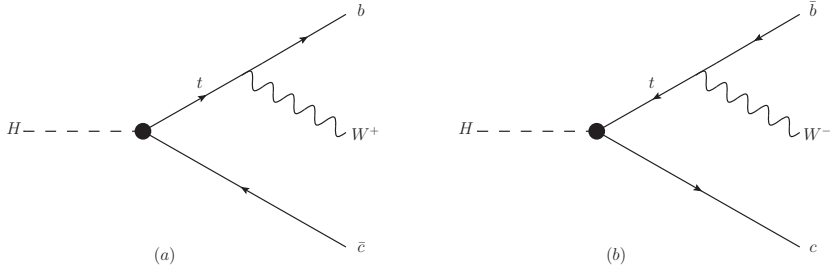}} \caption{Feynman diagrams of
the rare decay $H\rightarrow W^{+}b\bar{c}$.}\label{hwbc}
\end{figure}

\begin{figure}[htbp]
\scalebox{0.45}{\epsfig{file=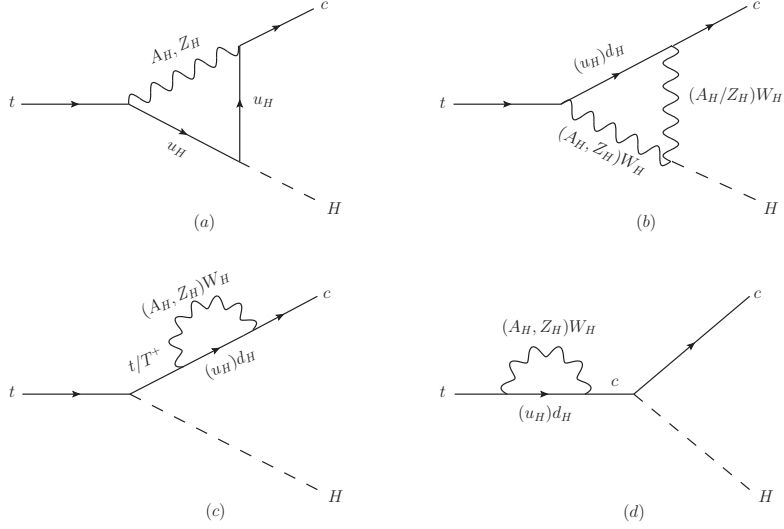}}\vspace{-0.5cm}\caption{Feynman
diagrams of the LHT one-loop correction to vertex $V_{tcH}$ in
unitary gauge.}\label{tch}
\end{figure}

The Feynman diagrams of the tree level $H\rightarrow W^{+}b\bar{c}$
and the rare decay $H\rightarrow W^{+}b\bar{c}$ are shown
respectively in Fig.\ref{hwbc2} and Fig.\ref{hwbc}, which includes
the $W^{+}$ and $W^{-}$ modes. The rare Higgs decay $H\rightarrow
Wbc$ is mediated by the same Yukawa coupling that leads to the
$t\rightarrow cH$ decay\cite{tchLHT}, so we show the Feynman
diagrams of the LHT one-loop correction to vertex $V_{tcH}$ in
unitary gauge in Fig.\ref{tch}, where the Goldstone bosons do not
appear. We can see that the flavor changing interactions between SM
quarks and mirror quarks are mediated by the heavy gauge bosons
$W_{H}^{\pm},Z_{H},A_{H}$. We find that dominant contribution to the
branching ratio of the decay $H\rightarrow Wbc$ is from the
interference between the Fig.\ref{hwbc2} and Fig.\ref{hwbc}.  Each
loop diagram is composed of some scalar loop functions \cite{loop
function}, which are calculated by using LOOPTOOLS\cite{loop tools}.


In our numerical calculations, we take the SM parameters as
follows\cite{parameters}
\begin{eqnarray}
\nonumber &&G_{F}=1.16637\times 10^{-5}\textrm{GeV}^{-2},
~\sin^{2}\theta_{W}=0.231,~\alpha_{e}=1/128,~m_{H}=125\textrm{GeV},\\
&&~~~~m_{c}=1.275\textrm{GeV},~m_{b}=4.18\textrm{GeV},~m_{t}=173.2\textrm{GeV},
~M_{W}=80.385\textrm{GeV}.
\end{eqnarray}

The LHT parameters related to our calculations are the scale $f$,
the mixing parameter $x_{L}$, the Yukawa couplings $\kappa_i$ of the
mirror quarks and the parameters in the matrices $V_{Hu},V_{Hd}$.
Due to the weak influence of the mixing parameter $x_{L}$, we take
$x_{L}=0.1$ for an example in our calculations. For the mirror quark
masses, we get $m_{u_{H}^{i}}=m_{d_{H}^{i}}$ at $\mathcal O(v/f)$
and further assume
\begin{equation}
m_{u_{H}^{1}}=m_{u_{H}^{2}}=m_{d_{H}^{1}}=m_{d_{H}^{2}}=M_{12}=\sqrt{2}\kappa_{12}f,~~
m_{u_{H}^{3}}=m_{d_{H}^{3}}=M_{3}=\sqrt{2}\kappa_{3}f.
\end{equation}
For the Yukawa couplings, the search for the mono-jet events at the
LHC Run-1\cite{k-LHC1} give the constraint $\kappa_i\geq 0.6$.
Considering the constraints in Ref.\cite{constraints}, we scan over
the free parameters $f$, $\kappa_{12}$ and $\kappa_{3}$ within the
following region
\begin{eqnarray*}
500{\rm GeV}\leq f\leq 2000{\rm GeV},~~0.6\leq \kappa_{12}\leq
3,~~0.6\leq \kappa_{3}\leq 3.
\end{eqnarray*}
For the parameters in the matrices $V_{Hu},V_{Hd}$, we follow
Ref.\cite{case} to consider the two scenarios as follows

\begin{itemize}
\item {Scenario I:}
$V_{Hd} = \textrm{I}$,$V_{Hu} = V^{\dag}_{CKM}$;
\item {Scenario II:}
 $s^d_{23}=\frac{1}{\sqrt{2}},
 ~s^d_{12}=s^d_{13}=0,
 ~\delta^d_{12}=\delta^d_{23}=\delta^d_{13}=0$.
\end{itemize}

\begin{figure}[htbp]
\begin{center}
\scalebox{0.4}{\epsfig{file=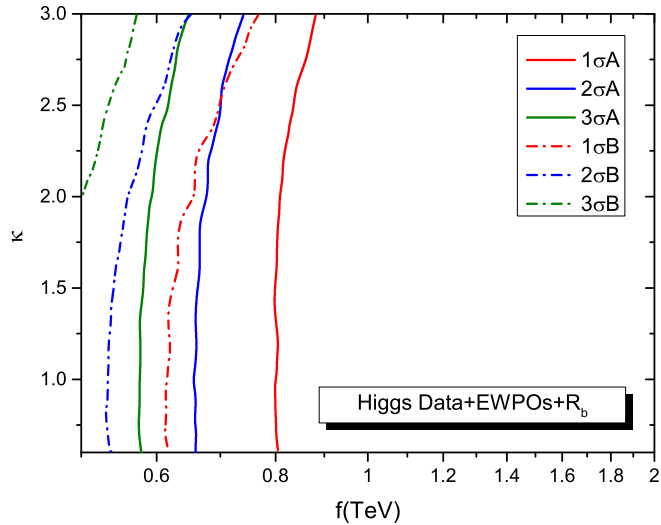}}\vspace{-0.5cm}
\caption{Excluded regions (above each contour) in the $\kappa\sim f$
plane of the LHT model for Case A and Case B, where the parameter
$R$ is marginalized over. The solid lines from right to left
respectively correspond to $1\sigma$, $2\sigma$ and $3\sigma$
exclusion limits for case A, and the dash lines correspond to the
case B.}\label{fkab}
\end{center}
\end{figure}

Furthermore, we will consider the constraint from the global fit of
the current Higgs data and the electroweak precision observables
(EWPOs)\cite{global fit}. In Fig.\ref{fkab}, we present the excluded
regions by the global fit of the Higgs data, EWPOs and $R_{b}$ in
the $\kappa\sim f$ plane of the LHT model for Case A and Case B,
where the parameter $R$ is marginalized over. In this global fit,
the three generation Yukawa couplings $\kappa_{i}$ are considered to
be degenerate, which will give a stronger constrain than the
nondegenerate case here.

\begin{figure}[htbp]
\begin{center}
\scalebox{0.4}{\epsfig{file=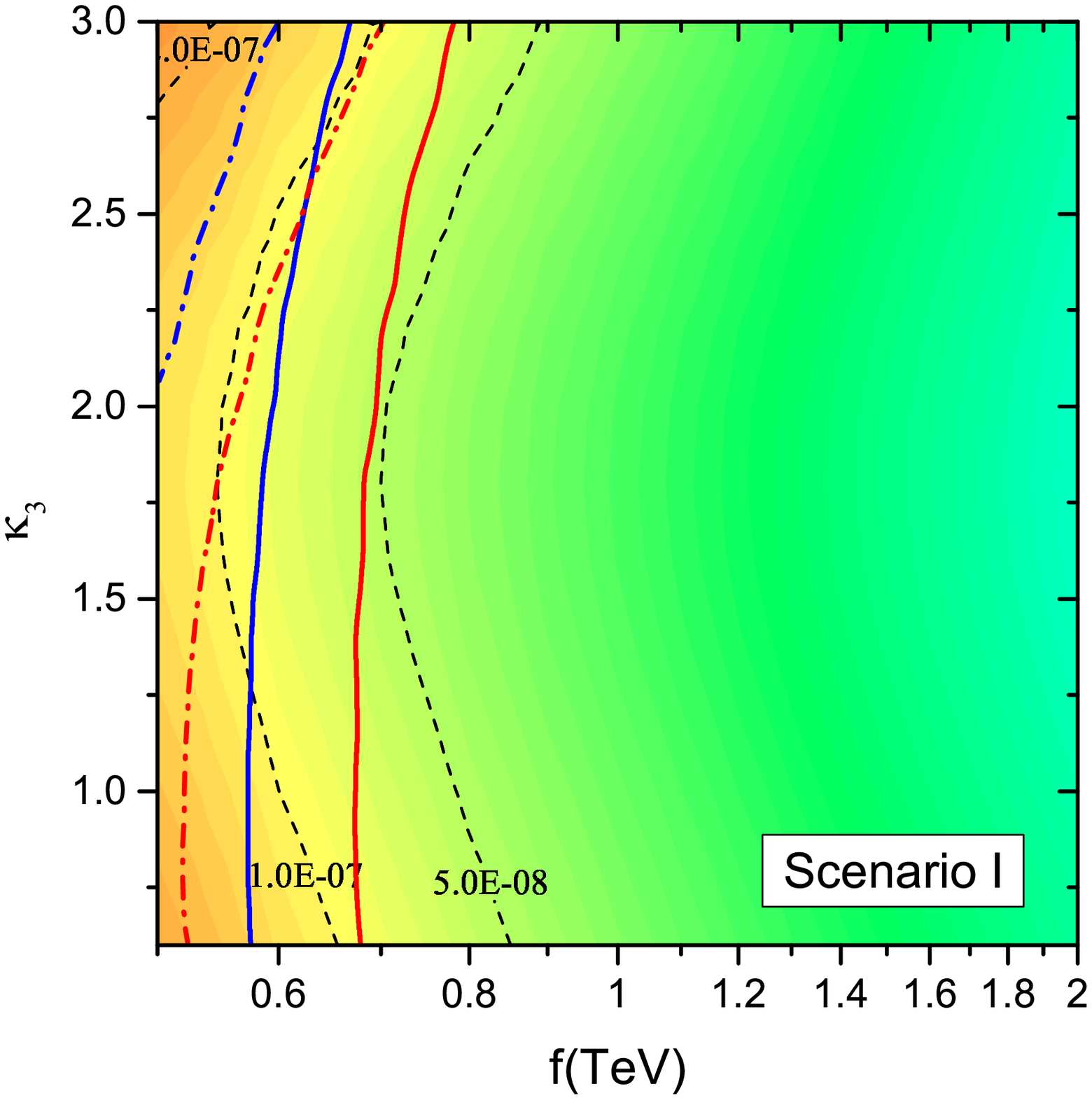}}\vspace{-0.5cm}\hspace{-0.cm}
\scalebox{0.4}{\epsfig{file=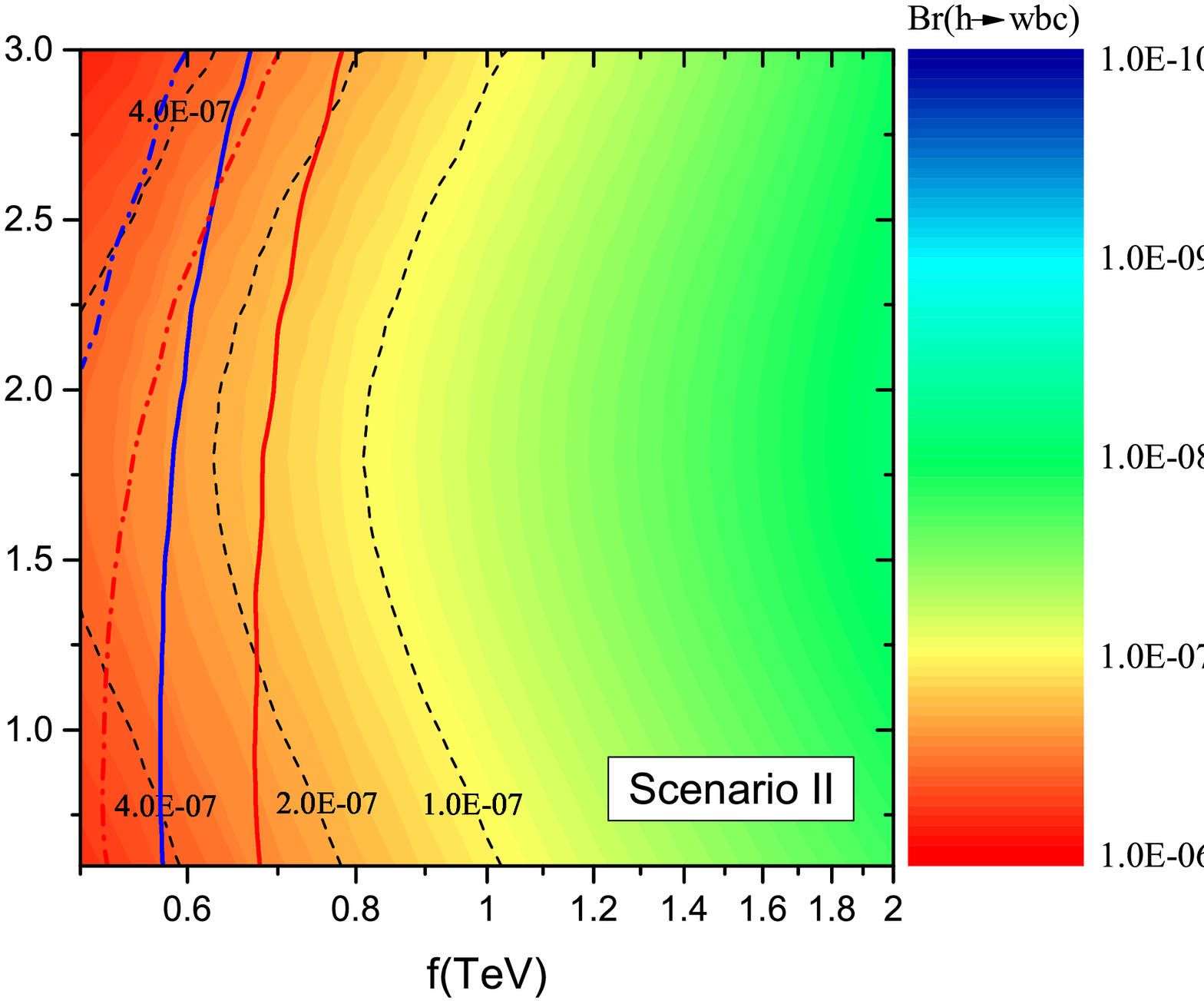}} \caption{Branching ratios of
$H\rightarrow Wbc$ in the $\kappa_{3}\sim f$ plane for two scenarios
with excluded regions of Case A and Case B, respectively. The red
lines and blue lines respectively correspond to $1\sigma$ and
$2\sigma$ exclusion limits as shown in
Fig.\ref{fkab}.}\label{fkcase12}
\end{center}
\end{figure}

In Fig.\ref{fkcase12}, we show the branching ratios of $H\rightarrow
Wbc$ in the $\kappa_{3}\sim f$ plane for two scenarios with excluded
regions of Case A and Case B, where the $W^{+}$ and $W^{-}$ modes
have been summed. From the left panel of Fig.4, we can see that the
branching ratio of $H\rightarrow Wbc$ in scenario I can reach
$1\times10^{-7}$ at $2\sigma$ level for Case A. This branching ratio
will become larger under the constrain of Case B. From the right
panel of Fig.\ref{fkcase12}, we can see that the branching ratio of
$H\rightarrow Wbc$ in scenario II can reach $4\times10^{-7}$ at
$2\sigma$ level, which is three even four times larger than that in
scenario I. Comparing the two scenarios, we can find that the
enhanced effects come from the large departures from the SM caused
by the mixing matrice in scenario II. From the two panels of
Fig.\ref{fkcase12}, we can see that the large branching ratios
mainly lie in the upper-left and lower-left corners of the contour
figures, where the scale $f$ is small and the Yukawa coupling
$\kappa_{3}$ is either too small or too large.

\begin{figure}[htbp]
\begin{center}
\scalebox{0.4}{\epsfig{file=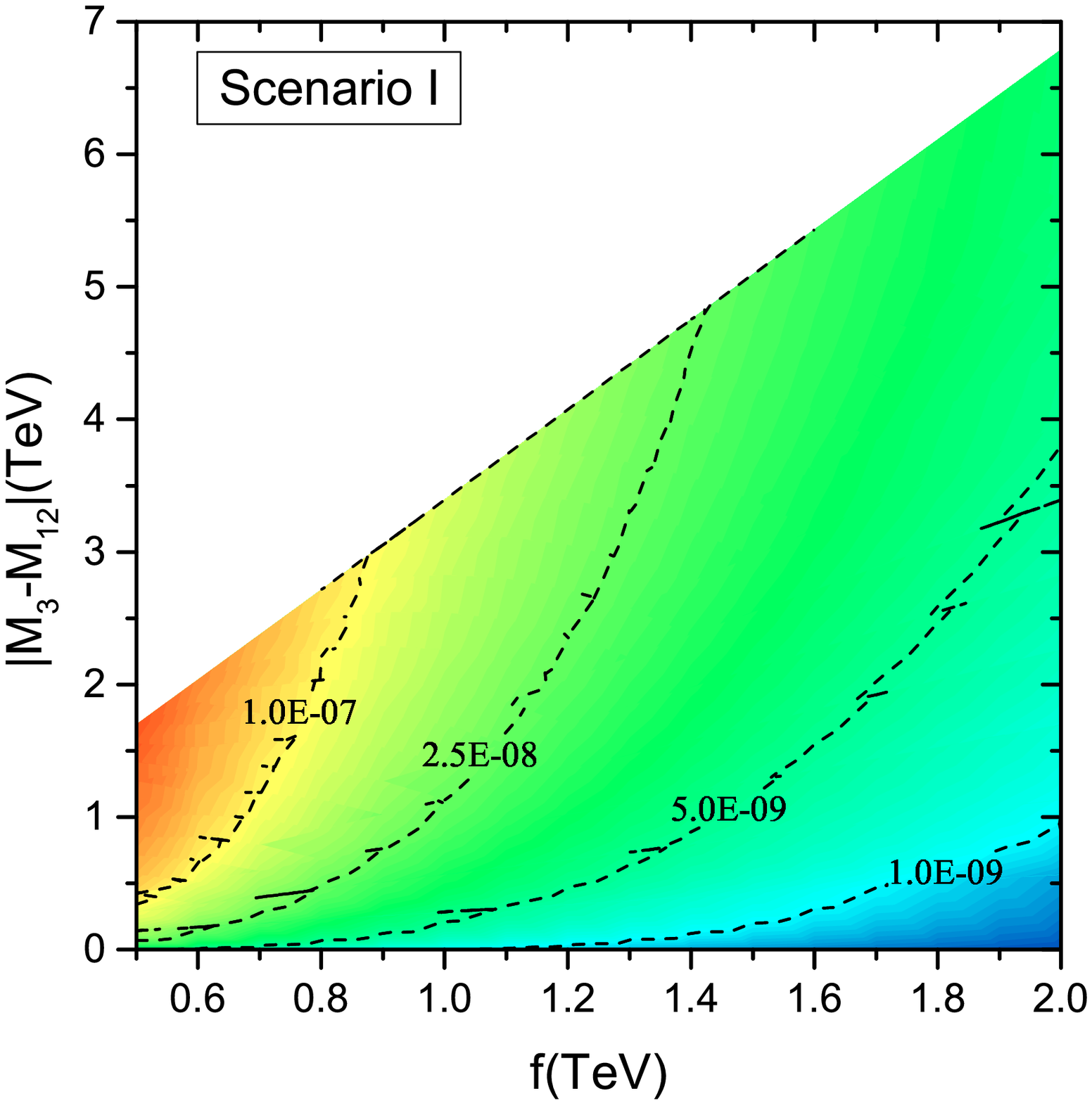}}\vspace{-0.5cm}\hspace{-0.cm}
\scalebox{0.4}{\epsfig{file=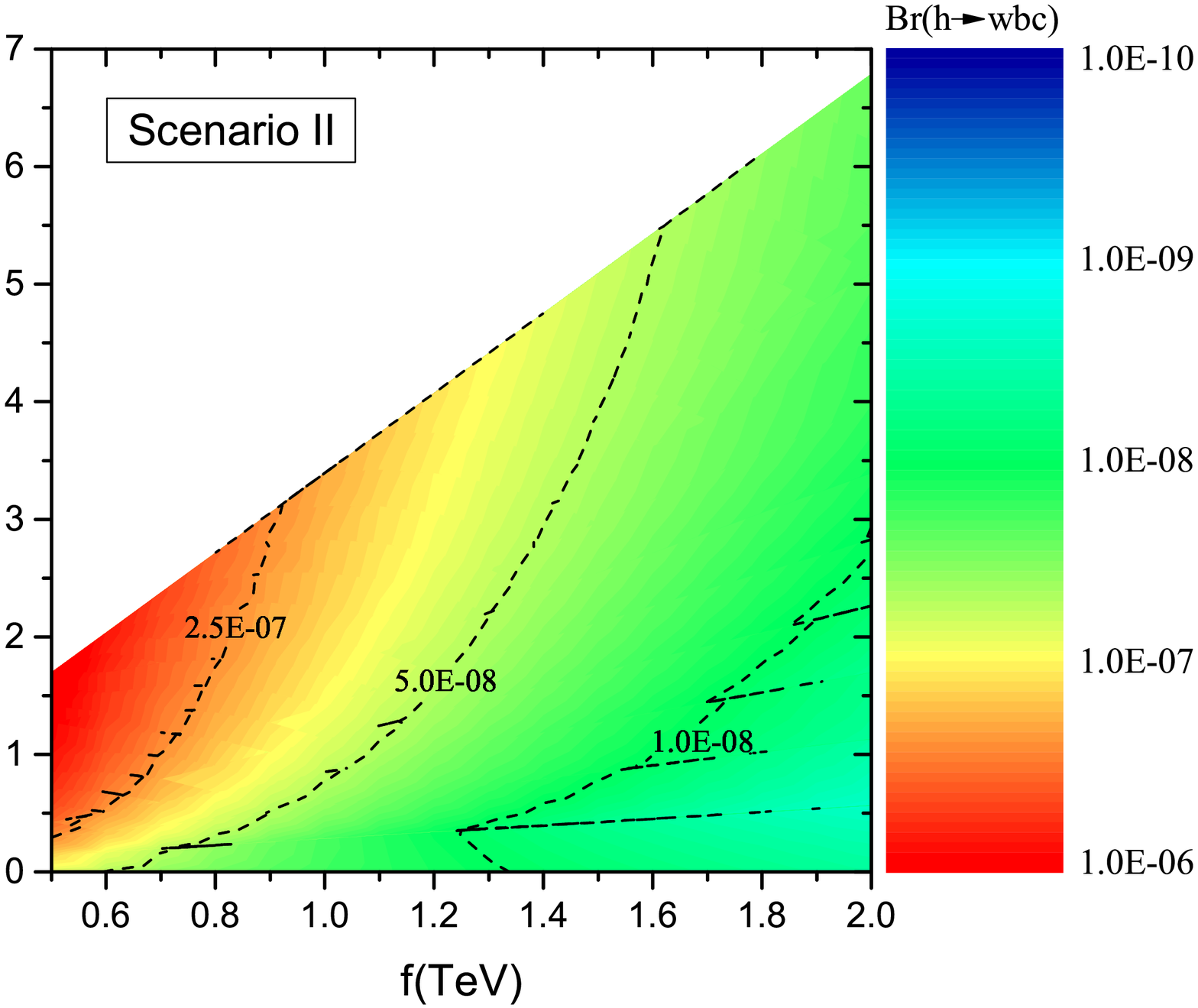}} \caption{Branching ratios of
$H\rightarrow Wbc$ in the $\mid M_{3}-M_{12}\mid\sim f$ plane for
two scenarios.}\label{case12}
\end{center}
\end{figure}

According to the Ref.\cite{case}, the branching ratio of
$t\rightarrow cH$ is enhanced by the mass splitting between the
three generation mirror quarks, the same thing will happen to the
branching ratios of $H\rightarrow Wbc$. In order to see this
dependence, we show the branching ratios of $H\rightarrow Wbc$ in
the $\mid M_{3}-M_{12}\mid\sim f$ plane for two scenarios in
Fig.\ref{case12}, respectively. We can see that the small branching
ratios correspond to the region that has small mass splitting $\mid
M_{3}-M_{12}\mid$ values. The largest branching ratios lie in the
upper-left corners of the contour figure with small $f$ and $\mid
M_{3}-M_{12}\mid$ of $1\sim2$ TeV rather than the regions that have
the largest $\mid M_{3}-M_{12}\mid$, that is because the branching
ratios are suppressed by the high scale $f$.

For the observability, the SM decay $H\rightarrow
WW^{\ast}\rightarrow Wbc$ is an important irreducible background
that will generate the same final state. Due to the off-shell top in
the signal decay $H\rightarrow t^{\ast}c\rightarrow Wbc$, we can use
the invariant masses cut $|M_{Wb}-m_{t}|> 20$ GeV to isolate the
signal. Besides, the $c$-jet in our signal comes from the Higgs
decay, which usually is harder than that in the SM background
$H\rightarrow WW^{\ast}\rightarrow Wbc$. Thus, we can use the high
transverse momentum $p_{T}^{c}$ cut to suppress the background.

Due to the same Yukawa couplings that lead to the $t\rightarrow cH$
decays, the decays $H\rightarrow t^{\ast}c\rightarrow Wbc$ can be
indirectly constrained by ATLAS and CMS searches\cite{LHC-tch}:
Br$(H\rightarrow t^{\ast}c\rightarrow Wbc)\leq 5.73\times 10^{-4}$,
where the $W^{+}$ and $W^{-}$ modes have been summed over. At the
LHC, the $t\bar{t}(\rightarrow WbWb)$ background is undoubtedly a
challenge, which will complicate the analysis for detecting the
decay $H\rightarrow t^{\ast}c\rightarrow Wbc$. Given this, the
linear collider with clean background may be an ideal place for
investigating this process, for example a future muon collider could
test the FCNC decay $t\rightarrow cH$ via Higgs decay $H\rightarrow
t^{\ast}c\rightarrow Wbc$ down to values of Br$(t\rightarrow cH)
\sim 5\times 10^{-3}$\cite{hwbc}.

\section{Conclusions}

\noindent

In this paper, we calculated rare Higgs three body decay
$H\rightarrow Wbc$ induced by top-Higgs FCNC coupling in the LHT
model. According to the parameters in the mixing matrices, we
considered two scenarios and found that the branching ratio for
$H\rightarrow Wbc$ can reach $\mathcal O(10^{-7})$ in the allowed
parameter space.

\section*{Acknowledgement}
This work is supported by the National Natural Science Foundation of
China (NNSFC) under grant Nos.11305049, 11405047, by the Startup
Foundation for Doctors of Henan Normal University under Grant
Nos.11112, qd15207 and the Education Department Foundation of Henan
Province(14A140010).

\end{CJK*}
\end{document}